\documentclass[12pt]{article}

\usepackage[margin=1in,bottom=0.95in,top=0.95in]{geometry}

\usepackage{indentfirst}
\usepackage{graphicx}
\usepackage{amsmath,amsthm,amssymb,verbatim,url}
\usepackage{xifthen}
\usepackage[dvipsnames]{xcolor}
\usepackage{wasysym}
\usepackage{hyperref}
\usepackage{xifthen} 
\hypersetup{colorlinks=true,urlcolor=blue}
\usepackage[affil-it]{authblk}

\newcommand{\bv}{\begin{verse}}
\newcommand{\ev}{\end{verse}}
\newcommand{\be}{\begin{equation}}
\newcommand{\ee}{\end{equation}}
\newcommand{\bea}{\begin{eqnarray}}
\newcommand{\eea}{\end{eqnarray}}
\newcommand{\bq}{\begin{quotation}}
\newcommand{\eq}{\end{quotation}}
\newcommand{\tr}{{\rm tr}}
\newcommand{\myurl}[2][]{\ifthenelse{\isempty{#1}}{\url{#2}}{\href{#1}{\tt #2}}}

\font\myfont=cmr12 at 18pt

\title{Quantum Dynamics Happens Only on Paper\\
{\myfont QBism's Account of Decoherence}}

\author[1]{John B. DeBrota}
\author[1,2]{Christopher A. Fuchs}
\author[1,3]{R\"udiger Schack}\footnotesize
\affil[1]{ Center for Quantum Information and Control, University of New Mexico\protect\\ Albuquerque,  NM 87131, USA}
\affil[2]{ Department of Physics, University of Massachusetts Boston\protect\\ 100 Morrissey Boulevard, Boston, 02155, MA, USA}
\affil[3]{ Department of Mathematics, Royal Holloway, University of London\protect\\ Egham, Surrey, TW20 0EX, UK}

\date{}

\begin{document}

\maketitle
\thispagestyle{empty}
\bigskip

\begin{abstract}
QBism has long recognized quantum states, POVM elements, Kraus operators, and even unitary operations to be cut from the same cloth: They express aspects of an agent's belief system concerning the consequences (for her) of actions she might take upon her external world. Such action-consequence pairs have conventionally been called ``quantum measurements.'' The calculus of quantum theory is then viewed as an empirically motivated {\it addition\/} to Bayesian decision theory when brought to this notion of measurement. This radical approach has allowed QBism to eliminate conceptual problems that plague other interpretations of quantum mechanics. However, one issue has remained elusive: If a QBist does not believe in the existence of an ontic (agent-independent) dynamical variable evolving over time, why would there be any constraints on her quantum-state assignment in the absence of performing a measurement? Why would she introduce unitary or open-system quantum dynamics at all? Here, we present a representation theorem based on van Fraassen's reflection principle to answer these questions. Simply put, an agent's assignment of quantum dynamics represents her belief that a measurement action she is contemplating would not change her current odds for future gambles.  A corollary to this approach is that one can make sense of ``open-system dynamics'' without the need to introduce an ``environment with a measurement record,'' as is common in decoherence accounts of quantum measurement. QBism's understanding instead rests more fundamentally on an agent's beliefs about the system of interest (not system plus environment) and her judgments about measurements she might perform on that system.  More broadly, this result establishes QBism's contention that measurement itself is the central concept of quantum theory and thus the framework upon which any future QBist ontology must hang.
\end{abstract}

\bigskip

\section{Introduction}

\label{OhWolfgangWolfgang}

\begin{flushright}
\baselineskip=13pt
\parbox{3.7in}{\baselineskip=13pt
Something only really happens during an observation \ldots\ as Bohr and Stern have finally convinced
me. Between the observations nothing at all happens, only time has \ldots\ progressed on the mathematical papers.\\
}
\\
--- Wolfgang Pauli, letter to Markus Fierz 1947
\end{flushright}

For nearly 100 years the central issue sustaining the quantum-interpretation debate has been that there seem to be two kinds of quantum-state evolution.  The usual story is that when a system is not observed, its quantum state evolves unitarily, whereas upon observation the state “collapses.”  The scare quotes are important.  This is because it is almost universally accepted that of the two unitary evolution is no mystery, whereas collapse is something no one understands.  This is the so-called quantum measurement problem.

If we were to trace the roots of this problem to a single source, it  would be that in conventional approaches, quantum states are treated as states of reality, without any fundamental need to speak of observers or measurements.  As a case in point, it is well known that John Bell declared that the term measurement ``should \ldots\ be banned altogether in quantum mechanics''~\cite[p.\ 209]{Bell2001}.  He prefaced this however with a sarcastic comment. ``Was the wavefunction of the world waiting to jump for thousands of millions of years until a single-celled living creature appeared \ldots\ [or] some better qualified system with a PhD?''  On display here is the tacit assumption that quantum states are states of reality with nothing {\it a priori\/} to do with the agent assigning them.

QBism \cite{Fuchs10a,Fuchs13a,Hero,AJP,Fuchs2017} in distinction is an interpretation of quantum theory that denies the existence of a measurement problem by insisting the theory should be understood along orthogonal lines to Bell's.  Quantum states according to QBism are not to be thought of as states of reality, but rather tools decision-making agents adopt to help choose which actions to take upon their external worlds.  The word ``measurement'' as commonly understood is thus a misnomer. In QBism the focus instead is on {\it actions\/} and {\it consequences}.  When an agent takes an action on an object, she confronts her external world with a moment of novelty it cannot anticipate; it responds with its own novelty in the form of a consequence for the agent. A quantum ``measurement outcome'' is born between the two---it represents something new in the universe, a moment of creation so to speak, mechanistically determined by nothing. That, QBism says, is what the probabilities in quantum theory are about: Agents' personal degrees of belief concerning the personal consequences of their potential actions.

In this light, QBism sees quantum theory as a formalism for providing agents with normative constraints which aid in shaping their personal beliefs, and thus gambles and decisions~\cite{Fuchs13a,DeBrota2021}.  It is a tool for checking whether one's various gambles are consistent or inconsistent.  These constraints are in addition to the ones given by personalist Bayesian probability theory~\cite{Bernardo1994}. This is because whereas Bayesianism is ``all-purpose'' for any possible world, quantum theory is tuned to the particular character of the world in which we live. The empiricist element of QBism is that if the world were different, the decision theory best adapted to the world would be different.

More too, this decision-theoretic conception becomes a rigorous demand on how to understand the other entities in the theory.  It implies that QBism cannot stop with quantum states being the sole bearers of a personalist judgment.  Complete consistency requires that QBism must say similar things for almost all the terms within the theory~\cite{Fuchs2002}: Measurement operators, Kraus operators, unitaries, and even Hamiltonians must be in the service of checking for consistency and inconsistency in an agent's mesh of beliefs.

This leads to something of a conundrum even from a QBist perspective however.\footnote{We acknowledge Harvey Brown~\cite{Brown2007} for being the first and only person to ever bring this up to us, and already in 2007 before the word QBism even existed!} We address the issue in this paper. The notion of measurement as an agent-centric notion is fundamental to QBism's whole enterprise.  But then as such, quantum dynamics ought to be a derived concept.  It is not enough to recognize quantum dynamics as a personal judgment and leave it at that.  Judgment of what?  At one level, the answer should be obvious: Dynamics is a judgment an agent makes for how she intends to update her quantum states over time. But then, why would she feel compelled to update her quantum states in the precise way the quantum formalism seems to mandate? Why are there any constraints at all on how an agent's state assignments should change over time when nothing is happening for her by way of an action on a system or a consequence arising from it?  {\it Nothing happens for the agent\/} employing the quantum formalism during the gap represented by dynamics; the agent receives no new experience from her external world.

Or, consider things from another perspective.  In the literature on open quantum systems and decoherence, one often sees descriptions such as this~\cite[p.\ 44]{Zurek1991}:  ``The environment can also monitor a system, and the only difference from a man-made apparatus is that the records maintained by the environment are nearly impossible to decipher.''  What records?  To QBism, this is anathema; it has no meaning.  The environment is not an agent that can employ the formalism of quantum theory for making better decisions.  The usual story as exemplified by Zurek's quote is an attempt to hide the structure of quantum theory in a classical worldview of an objective agent-independent state evolving deterministically.

Thus QBism is behooved not only to derive quantum dynamics as a constraint on how an agent's state assignments should change with time, but also to tell a story of open-system dynamics and decoherence that never invokes an environment in any essential way---certainly not one that ever invokes unobserved measurement records.  There can be nothing conceptually like a Maxwellian demon living in the environment making records in QBism's story.

Said in this way, the problem starts to take the shape of something encountered long ago with the issue of the quantum de Finetti theorem~\cite{Caves2002b,Fuchs2004a,Fuchs2004b}.  There, what was sought was a QBist understanding of quantum-state tomography.  In a conventional description of tomography, a preparation device repeatedly prepares systems in some fixed but unspecified ``unknown quantum state.'' The systems are then presented to an observer whose task is to estimate the unknown state after performing a large number of measurements, system by system. This description makes no sense from a QBist perspective: In QBism, any time a quantum state is invoked it must represent some agent's beliefs.  Consequently it must be known to that agent.  But barring the fiction of a Maxwellian demon sitting inside the preparation device making the ``unknown state'' as his quantum-state assignment, there are no other agents around but the one outside the device making the measurements.

What was called for was a way of thinking about the set-up of quantum tomography that involves only the single agent performing the measurements and the quantum-state assignments---known to her of course---that she actually makes.  This was where the quantum de Finetti representation theorem stepped in to save the day.  Through it one could see that, so long as the agent makes an initial quantum-state assignment to the collective of systems emitted from the device which has a certain exchangeability property,  one can {\it represent\/} that assignment as a mixture of unknown quantum states. The real thing is the single state assignment, all the rest is artifice, the result of a representation theorem.

Here we do something similar for quantum dynamics.  We prove that quantum dynamics can be seen as a statement recovered from a representation theorem, the sole premise of which has to do with the state changes an agent expects to make should she perform a measurement at some stage in the future.  The representation theorem thus gives a way of thinking of the notion of measurement and its associated notion of state change as logically prior to dynamics.  Our approach accounts for decoherence without invoking an unobserved measurement record, just as the de Finetti theorem accounts for quantum tomography without invoking an unknown quantum state.

To frame the setting of our considerations, it is worth repeating the observation that motivates us:  In the absence of a measurement action, a more fundamental ``dynamics-less'' QBism does not automatically provide guidance for how a quantum-state assignment should change over time.  The only fundamental constraints on state changes in QBism ought to arise from measurement.

This does not mean however that a QBist is prohibited from changing her state for a physical system unless she acts on it. There are any number of essentially nonquantum reasons an agent might change her state assignment before an upcoming measurement. For instance, between her last measurement and the next, she might do additional calculations that she didn't have time to do previously. Or she might revise her prior assumptions because they start to seem wildly unreasonable. Or maybe she simply forgot some other considerations she would like to take into account.  These things are allowed in the most liberal of personalist Bayesian frameworks~\cite{Hacking1967}.  They are a simple consequence of the conception of probability assignments as tools for decision making, not facts of nature.

Yet, there is a difference when it comes to quantum measurement.  What is distinct about state changes that arise from measurement is that they derive from new physical experience---they arise from the agent's action on her external world and are not sprung from her thought, regret, or recalculation.  For this case, the guidance of quantum theory is unambiguous:  There must exist a set of Kraus operators to take the initial state assignment to an updated state~\cite{Kraus1983}.\footnote{Nielsen \& Chuang~\cite{Nielsen2010} call them ``measurement operators.''  We call them Kraus operators because it is common terminology based on his influential book {\sl States, Effects and Operations:\ Fundamental Notions of Quantum Theory}, but we note that the concept first appeared 14 years earlier in the work of Hellwig \& Kraus~\cite{Hellwig1969,Hellwig1970}. Also see Footnote \ref{Delilah} for a certain qualification of this.}  In other words, the form of the state change rule is part of quantum theory's normative structure: In this paper, we take that idea to be fundamental and elaborate on it in Section~\ref{QBismSection}.\footnote{See Ref.~\cite{Fuchs2023} for a more extensive account of QBism's understanding of quantum theory's normativity than we have room for here.}

The thing we do next is show that this normative requirement from quantum theory combined with van Fraassen's reflection principle, a little-known normative requirement from personalist Bayesian probability theory~\cite{vanFraassen1984,Shafer1983,Goldstein1983}, leads to the notion of quantum dynamics.  Loosely speaking, our result is this:  When an agent makes the judgment of a trace-preserving completely positive map connecting states through time (i.e., an instance of quantum dynamics), she is really expressing her underlying belief that there is a measurement action she could take in the future that would make no difference for {\it how she should gamble now\/} on a still farther future. Conceptually, quantum dynamics is not just about an initial time $t_0$ and a final time $t_2$; it is about something that might happen for the agent in between (at a time $t_1$ say) which she deems to be of no consequence for her $t_0$ gambles.

\begin{figure}[t]
    \centering
    \includegraphics[width=\textwidth]{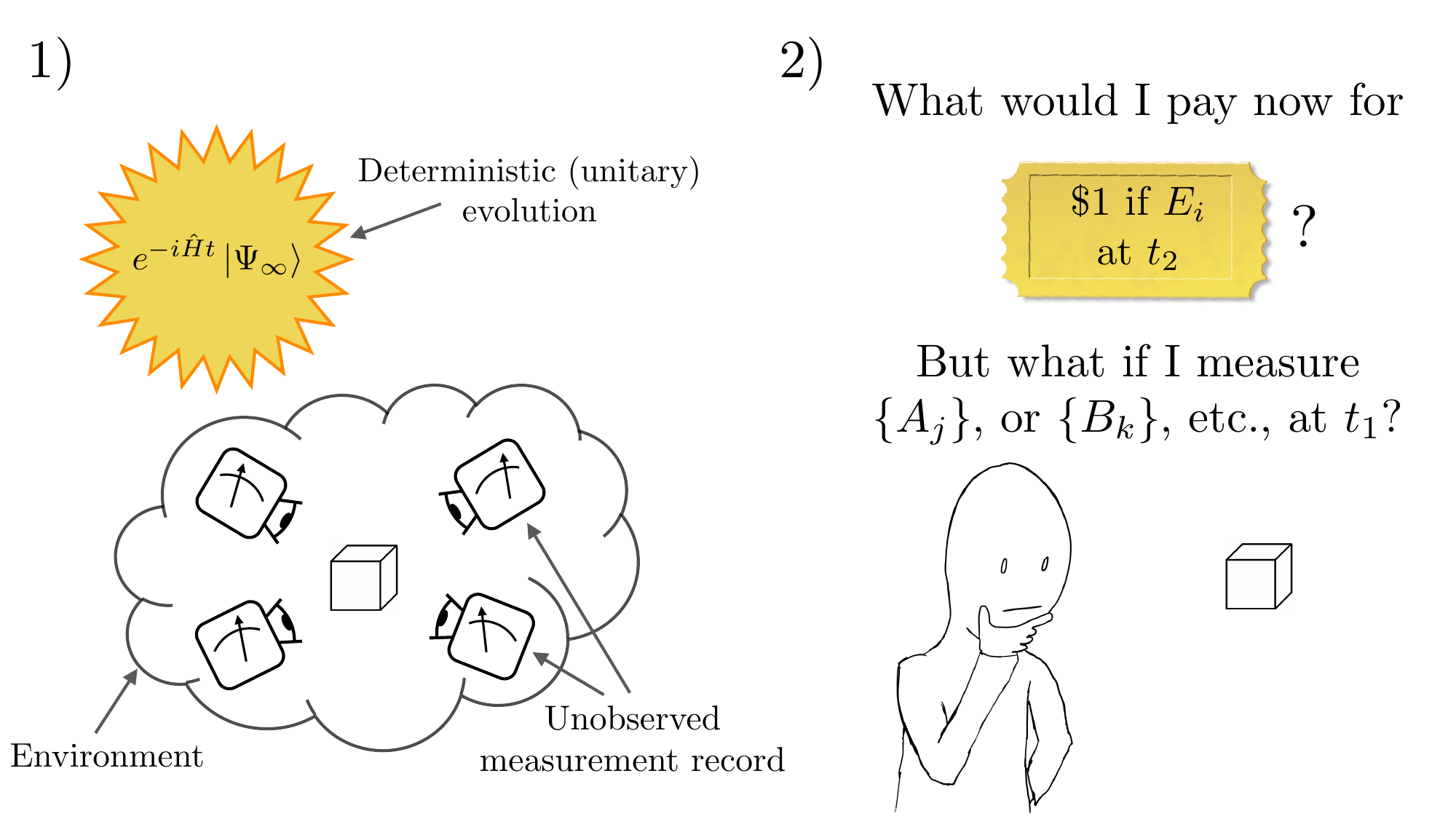}
    \caption{1) The usual story is that quantum dynamics and decoherence are physical processes that happen to a system. If the system is isolated, the quantum state of the system evolves unitarily; if the system is not isolated, decoherence results from the system's unitary interaction with its environment. This story is plagued by the notorious measurement problem. 2) In QBism, quantum dynamics is like futures pricing. An agent is attempting to gauge what she would regard as a fair price for a contract that guarantees a payout of \$1 if a measurement action she will take at maturity (a future moment in time) results in a given outcome. To get a handle on this, she tries to identify measurements she could take between now and maturity that would not change what she regards as a fair price now. Quantum dynamics {\it happens\/} only on the futures trader's spreadsheet.}
    \label{fig:tickets}
\end{figure}

Perhaps an analogy to futures trading would be useful. (See Fig.\ \ref{fig:tickets}\@.) Suppose the futures in question are lottery tickets that will pay \$1 at maturity---i.e., at a time $t_2$---if a certain event occurs {\it at that time}.  What should a trader consider a fair price for buying or selling them at $t_0$, knowing full well that she will still be in the business of buying and selling them all the way until $t_2$?  If the trader could foresee {\it the actual future\/} then she should adjust the $t_0$ price to be precisely the $t_2$ price---else she would expose herself to the possibility of a sure loss. But such knowledge of the future is beyond her grasp. The best she can do is see to it that the price she holds at $t_0$ and the {\it expectation value\/} of the prices she might hold at $t_2$ are in accordance with each other. Neither variable is the more primary from a logical point of view; it is just the identity that must be insured.  This is what van Fraassen's reflection principle is about~\cite{Fuchs2012a}.

Notice though that to maintain van Fraassen's principle, both ends of the pricing must be resilient to what the agent expects to be doing in between.  For instance, if the trader expects to only consult publicly available information for real-time pricing adjustments between $t_0$ and $t_2$, then the natural outcome for the $t_0$ pricing will be a probabilistic mixture of the hypothesized real-time $t_2$ pricings. On the other hand, if the trader wants to hold fast to a predetermined $t_0$ pricing at all costs, then perhaps a drastic action such as leaking insider information will be necessary.\footnote{Or to mix metaphors, perhaps tie herself to a mast as in the tale of Ulysses and the sirens~\cite{Fuchs2012a,vanFraassen1995}.}  In either case, the transactions and expectations refer to the agent's standpoint at $t_0$.

What especially interests us for the quantum case is the first of these two examples, where the idea of consulting publicly available information will be replaced with a quantum measurement taking place at some time between $t_0$ and $t_2$. In general, it will matter to the trader's (or agent's) $t_0$ pricings whether or not she intends to carry out that in-between measurement. This provokes the following idea. Suppose there is an in-between measurement for which the trader judges that it would not matter for her $t_0$ pricings whether she chooses to actually carry it out or not. In the case that she does not intend to carry out the measurement, she will then expect to hold fast to the $t_0$ pricing when the time $t_2$ genuinely arrives. This is quantum dynamics, and it all happens at $t_0$.

With this outlook as our aim, the structure of our paper is as follows. In Section~\ref{QBismSection}, we give a barebones review of the parts of QBism necessary for our argument.  In Section~\ref{ReflectionSection}, we explain the setting of van Fraassen's reflection principle and then combine it with the quantum formalism in terms of Kraus state changes~\cite{Kraus1971}.  Finally, in Section~\ref{DynamicsSection}, we derive the main result of our paper---the representation theorem for dynamics in terms of a more fundamental ``dynamics-less'' QBism.  Thereafter, we apply the new conception of dynamics to examples in continuous quantum dynamics and decoherence in  Sections~\ref{ContinuousDynamicsSection} and \ref{DecoherenceSection}.  We conclude in Section~\ref{ConclusionsSection} by reiterating the importance of this work for establishing a more complete understanding of QBism.

First though, a few words are in order concerning the still deeper historical setting for this effort.  One of us has noted in several publications how the program of QBism descended from the thinking of John Archibald Wheeler in the 1970s and early 1980s~\cite{Fuchs2017,Fuchs2012b,Samizdat2,Fuchs2016}.  Prompted by the crisis he saw in the singularity theorems of general relativity, Wheeler turned to the quantum for guidance as to what the ultimate stuff of the world might be.  Somewhere in it all, it came to him that the stuff might be drawn from a category that subsumes quantum measurement as a particular instantiation when physicists are about.  He called the larger category ``acts of observer-participancy’’ in an attempt to de-anthropocentrize the notion, though to limited success and consistency.  Still the sentiment was there.  Wheeler would say things like this~\cite{Wheeler82c},
\begin{quote}
It is difficult to escape asking a challenging question. Is the entirety of existence, rather than being built on particles or fields of force or multidimensional geometry, built upon billions upon billions of elementary quantum phenomena, those elementary acts of ``observer-participancy,'' those most ethereal of all the entities that have been forced upon us by the progress of science?
\end{quote}
and his students would take note.  Indeed he inspired a few of us to take the question seriously.  Consider William K. Wootters in this little known work~\cite{Wootters1994}:
\begin{quotation}
From a pragmatic point of view, quantum mechanics can be thought of as a theory that predicts probabilities of the outcomes of certain measurements given certain prior measurements. One does not in principle have to talk about what goes on between measurements. However, the theory suggests strongly that something does go on between measurements, namely, a unitary evolution (either of the state vector or of the operators). It is very natural that many of us have come to take this unitary evolution not only as fundamental, but also as inviolable. \ldots\

However, this view is not the only one possible. According to the Copenhagen interpretation, measurements are not like other processes but have a special status. They do not correspond to reversible unitary transformations but rather to irreversible projections.\ldots

The emergence of a definite outcome of a quantum measurement, if one regards it as a fundamental process, is the only thing in fundamental physics that is not reversible. It is therefore a natural place to look for the origin of time asymmetry. \ldots

Despite [the objections outlined] I wish to explore briefly in this paper a line of thinking based on the view just presented, to see where it might lead. \ldots\ [O]ne hopes to obtain, from an argument that assumes nothing about Hilbert space or unitary transformations, a relationship among probabilities that is most conveniently {\it summarized\/} in terms of Hilbert space and unitary transformations. Probabilities come logically before, not after, state vectors. In particular, the existence of ``measurement events,'' the probabilities of whose outcomes are to be computed, is taken here to be fundamental.
\end{quotation}
Our current paper should be seen as a continuation of this tradition, though with a key difference in mindset from what one usually finds in quantum foundations discussions.  QBism has never gone straight for an ontology.  Instead, as was remarked in Ref.\ \cite{Fuchs2023}, ``QBism's tack was to ask over and over, what is it about the world that makes us well-advised to use the calculus of quantum mechanics for structuring our [decision theoretic] probabilities? Said this way, it became clear that the pertinent way to move forward was to get the `epistemics' of the theory right before anything else:  Getting reality right would follow for those who had patience enough to pass the marshmallow test~\cite{Mischel2014}.'' We hope the reader will understand the purpose of our paper in this light.

\section{QBism}
\label{QBismSection}

For QBism, quantum mechanics is about taking actions. At the heart of QBism is the notion of a quantum
measurement, an action an agent takes on her external world with consequences she generally cannot predict with certainty.\footnote{In other literature we have called these consequences ``experiences''~\cite{DeBrota2020} to emphasize that they should be understood in contrast to ``facts for everyone''~\cite{Brukner2020}. Perhaps the hint of this idea was first presented in Ref.\ \cite{Fuchs2007}.  For the the most current development of the idea in the philosophical context of phenomenology, see several papers in Ref.\ \cite{Berghofer2023}.}  When the consequences of the actions matter to the agent in the sense that they are worth her while to gamble upon them, she is well advised to use the quantum formalism to guide her decisions regarding which actions to take.

In this way, quantum mechanics should be viewed as a decision theory which functions as an addition to standard personalist probability theory. Assume our agent plans to make an $n$-outcome measurement on some quantum system. According to QBism, the measurement outcomes, labeled $i\in\{1,\ldots,n\}$, are personal to the agent---they need not be ``objectively available for anyone's inspection,'' as Wolfgang Pauli once wrote~\cite{Fuchs10a}. Furthermore, the quantum state $\rho$ the agent assigns to the
system, the POVM elements $E_i$ she associates with her outcomes, and the
corresponding probabilities $p_i$ she assigns to the latter, are also not objectively given, but
personal to the agent and reflect her beliefs.

Within the quantum formalism, the Born rule
\begin{equation}
p_i=\tr\,\rho E_i
\label{eq:born}
\end{equation}
links $\rho$, $E_i$ and $p_i$, all of which represent the agent's
beliefs. QBism regards the Born rule not as a rule to set $p_i$ once $\rho$ and
$E_i$ are given, but as a consistency criterion. If a given $p_i$, $\rho$, and $E_i$ satisfy Eq.\ (\ref{eq:born}) they are consistent, else they are not.  From this point of view the Born rule is a flagging device: It has three independent inputs and a single output, either consistent or inconsistent.  However note that when the Born rule flags inconsistency, it gives the agent no
prescription for how to resolve it.

Nonetheless, the Born rule is an extraordinarily powerful constraint on the agent's decision
making. For instance, the design of a quantum algorithm can be viewed as the
task of maximizing the agent's probabilities that an action she takes---in this
case carrying out a measurement at an appropriate time---will lead to a desirable consequence, e.g.,
that its result will reveal one factor of a composite number.

Something is missing from our story, however. Often, as
a consequence of a measurement action, an agent will update her probabilities
and state assignments regarding subsequent measurement actions. Let us thus
assume that our agent is considering a further measurement action---perhaps some $\{F_j\}$ yet to be determined---immediately following the $\{E_i\}$ measurement. Assume that the agent adopts
the following update rule: She assigns states $\rho_1,\ldots,\rho_n$ and makes
the commitment to use the state $\rho_i$ conditional on getting outcome $i$ of the set $\{E_i\}$. In other words, she commits to using $\rho_i$ to inform her subsequent decisions if she gets outcome $i$.

For handling this situation, the quantum formalism provides further consistency requirements on the agent's beliefs. In a later paper~\cite{DeBrota2024}, we will take up the full palette of possibilities consistent with QBism's embrace of genuinely personal quantum states.\footnote{\label{Delilah} These will include the linear but not-completely-positive maps defined on restricted domains of quantum state space advocated in Refs.\ \cite{Pechukas1994,Shaji2005,Joseph2018} and a larger notion even still.} For our current considerations though, it is enough to show how things work out in the usual setting of quantum information theory---as for instance exhibited in Nielsen \& Chuang's textbook~\cite{Nielsen2010}---where one makes use of a simple generalization of the well-known von Neumann and L\"uders rules for state updates.  Most often one sees it expressed in the simplified ``efficient measurement'' case~\cite{FuchsJacobs} as: There must be a set of so-called Kraus operators $A_i$ such that
\begin{equation}
E_i = A_i^\dagger A_i
\label{BoogalooSituation}
\end{equation}
and
\begin{equation}
\rho_i = \frac{1}{p_i}A_i \rho A_i^\dagger\;.
\label{eq:kraus}
\end{equation}
More generally, a further index $k$ may be required on the Kraus operators $A_{ik}$. When this is so, the appropriate connections become
\be
E_i = \sum_k A_{ik}^\dagger A_{ik}\;,
\label{ChurchBells}
\ee
and the state-change rule becomes
\be
\rho_i=\frac{1}{p_i}\sum_k A_{ik}\rho A_{ik}^\dagger
\label{WillBeRinging}
\ee
for all outcomes $i$.\footnote{Kraus called such a state change a ``selective operation'' and the associated $E_i$ an effect~\cite{Kraus1983}. We eschew the term ``operation'' in this paper because it suggests that something is changed in the world instead of in the agent's gambling attitudes.} In much of our upcoming discussion we will focus on the efficient-measurement case of Eqs.\ (\ref{BoogalooSituation}) and (\ref{eq:kraus}) for simplicity.  However, everything generalizes appropriately; so there is no worry of incompleteness.

As with the Born rule, QBism interprets the Kraus rule as a tool for an agent to detect inconsistencies in her beliefs. It is not a rule for how to find the ``right'' post-measurement state.  If given sets of $\{\rho_i\}$, $\{p_i\}$, and $\{A_{i}\}$, along with $\rho$, all satisfy Eq.\ (\ref{eq:kraus}), they are consistent. Else they are not, and the agent is well-advised to make some adjustments in her various terms until consistency can be met.

\subsection{An Apparatus-Free Dynamics-less Account of the Kraus Rules}

At this point, a special discussion is warranted as there is a common preconception that the very reason an agent uses the Kraus rules is because there is a dynamical model underlying the measurement in the first place.  The ``misguidedness'' of this perception is something we will elaborate on at length on in a follow-on paper~\cite{DeBrota2024}, but the crucial point for our current account is that the preconception is something one can either take or leave.  QBists leave it.

To build the setting, let us trace a bit of the history starting with the original Hellwig \& Kraus papers~\cite{Hellwig1969,Hellwig1970}. Their interest was in refining the mathematics of the most general quantum measurement model.  They described it thus.  First a system in a state $\rho$ interacts with an apparatus prepared in an independent (but standardized) state $\sigma$. Assuming the two systems are sufficiently isolated from the rest of the universe, this leads to a unitary operator evolution:
\be
\label{MothInTheNight}
\rho\otimes\sigma\quad \longrightarrow \quad U(\rho\otimes\sigma)U^\dagger\;.
\ee
Thereafter one measures the apparatus {\it alone\/} via a textbook von Neumann scheme.  This projects the quantum state of the whole onto a subspace identified with an eigenprojector $\Pi_i$ of some Hermitian observable.  Doing such causes a collapse of the joint quantum state according to:
\be
U(\rho\otimes\sigma)U^\dagger \quad \longrightarrow \quad \frac{1}{p_i}(I\otimes\Pi_i)\Big(U(\rho\otimes\sigma)U^\dagger\Big)(I\otimes\Pi_i)\;.
\ee
Finally, the apparatus is discarded from consideration via the mathematical operation of a partial trace.  With this final step, one arrives at a new quantum state $\rho_i$ for the system.
 
Notably, the description above insinuates that dynamics is indeed a crucial ingredient in the transition $\rho\,\rightarrow\,\rho_i\,$.  This is unacceptable for our current considerations. Yet, there is a golden nugget for QBism in the mathematics of these papers. Perhaps to their own surprise (and that is why they felt it worth recording in a paper), Hellwig \& Kraus showed in the end that the state-change process can be distilled into an expression {\it involving operators acting on the system's Hilbert space alone, not on the joint space}.  Thus, they arrive at a description of state change of the form Eq.~(\ref{WillBeRinging}), with the underlying origin story no longer immediately visible.  This was useful in a number of ways, not least of which was in making a connection to Stinespring's notion of completely positive maps on operator algebras~\cite{Stinespring1955}.  The state changes that occur under the most general notion of a quantum measurement are (linear) completely positive maps, followed by renormalization.

This is the conception we aim to turn on its head in this paper.  The key insight comes from the fact that QBism does not need the Hellwig \& Kraus origin story. This follows because of a mathematical insight of M.-D. Choi in 1975~\cite{Choi1975}.\footnote{This is a wonderful example of how a purely mathematical result---in this case one published in the journal {\it Linear Algebra and Its Applications}---can lead to significant philosophical insights.}  Through it, QBism can give a consistent scheme for updating quantum states full stop---no need to invoke dynamics at all.

A small aside first.  What, after all, is so bad about the original Hellwig \& Kraus origin story?  The answer lies in a number of reasons described in this paper, but let us emphasize one particularly pertinent one here.  Hellwig \& Kraus separate out the apparatus as a distinct entity from the agent. As emphasized especially in Refs.\ \cite{Fuchs2017,Fuchs2023} however, in QBism the measurement apparatus must be understood as a part of the agent herself---not as a separate part of the external world. Therefore in writing down a state $\sigma$ in Eq.~(\ref{MothInTheNight}), the agent would be making a quantum-state assignment to herself, which has no sense in QBism.  See e.g.~\cite[Sec.\ 2.3]{Fuchs2023} for more details.

So, the focus for QBism really must be on a consistent updating rule without any mechanical picture behind it:  New experience needs to combine with an old quantum-state assignment to make for a new quantum-state assignment. This is the requirement for our search.  By a general argument beyond the scope of the present paper (though it will appear in Ref.\ \cite{DeBrota2024}), one can deploy the reflection principle of Section \ref{ReflectionSection} still more extensively to show that for each outcome $i$ of a POVM $\{E_i\}$, there must be a linear map $\Phi_i$ such that
\be
\rho_i = \frac{1}{p_i}\Phi_i(\rho)\;.
\label{ParitallyThere}
\ee
As before, $p_i$ denotes the agent's subjective probability for the outcome $i$ as given by the Born rule (\ref{eq:born}).

This has a surprisingly simple consequence that follows from the mathematics of tensor products, without any underlying imagery such as that used by Hellwig \& Kraus. We need only make use of the dry mathematical facts that $(A\otimes B)(C\otimes D)=AC\otimes BD$ and $\tr A\otimes B=\tr A\, \tr B$ to find a pretty re-expression of the Born rule when applied to two consecutive measurements.

Consider again the scenario described previously of a measurement $\{F_j\}$ performed immediately after a $\{E_i\}$ measurement where an outcome $i$ was obtained.  Let us denote by $p(j|i)$ the conditional probability of getting $F_j$ after getting $E_i$.  Introducing an orthonormal basis $|k\rangle$ on the system's Hilbert space, the initial quantum state $\rho$ becomes
\be
\rho = \sum_{kl} \langle k|\rho|l\rangle\,|k\rangle\langle l|\;.
\ee
Using this, we have
\bea
p(j|i) &=& \frac{1}{p_i}\,\tr\Big(\Phi_i(\rho)F_j\Big)
\\
&=& \frac{1}{p_i}\,\tr\left[\left(\sum_{kl}\langle k|\rho|l\rangle\, \Phi_i\big(|k\rangle\langle l|\big)\right)\! F_j\right]
\\
&=&
\frac{1}{p_i}\, \sum_{kl} \langle k|\rho|l\rangle\,\tr\Big( \Phi_i\big(|k\rangle\langle l|\big) F_j\Big)
\\
&=&
\frac{1}{p_i}\, \sum_{kl} \tr\Big(|l\rangle\langle k|\rho\Big)\tr\Big( \Phi_i\big(|k\rangle\langle l|\big) F_j\Big)
\\
&=&
\frac{1}{p_i}\, \sum_{kl} \tr\Big(|k\rangle\langle l|\rho^{\rm T}\Big)\tr\Big( \Phi_i\big(|k\rangle\langle l|\big) F_j\Big)
\\
&=&
\frac{1}{p_i}\,\tr\!\left[\left(\sum_{kl}|k\rangle\langle l|\otimes \Phi_i\big(|k\rangle\langle l|\big)\right)\!\Big(\rho^{\rm T}\otimes F_j\Big) \right],
\eea
where $\rho^{\rm T}$ is the transpose of $\rho$ as expressed in the given basis.  This expression can be put into a more compact form by introducing a maximally entangled vector across the two artifice Hilbert spaces:
\be
|\Gamma\rangle=\sum_k |k\rangle|k\rangle\;.
\ee
With that,
\be
\sum_{kl}|k\rangle\langle l|\otimes \Phi_i\big(|k\rangle\langle l|\big)\,=\, I\otimes\Phi_i\big(|\Gamma\rangle\langle \Gamma|\big)\;,
\ee
and we get as our final result
\bea
p(j|i) &=& \frac{1}{p_i}\,\tr\Big(\Phi_i(\rho)F_j\Big)
\\
&=& \frac{1}{p_i}\,\tr\Big[\Big(I\otimes\Phi_i\big(|\Gamma\rangle\langle \Gamma|\big)\Big)\Big(\rho^{\rm T}\otimes F_j\Big)\Big]\;,
\label{ChoiReady}
\eea
two very distinct expressions for thinking about how to calculate the probabilities.

Looking at this formulation, one should ask: Is the extra Hilbert space in Eq.\ (\ref{ChoiReady}) {\it really\/} there as an ancilla, apparatus, or environment?  Of course not!  It was all just a mathematical trick.  Nonetheless, it is a trick that sets us up for Choi's profound theorem.  In this form, we can now see how to bring our partially-formed update rule in Eq.\ (\ref{ParitallyThere}) to finalization.

Choi's theorem states that $\Phi_i$ is a completely positive map---i.e., one that gives rise to an expression of the form Eq.\ (\ref{WillBeRinging}) for the $\rho_i$---if and only if the ``Choi matrix'' $I\otimes\Phi_i\big(|\Gamma\rangle\langle \Gamma|\big)$ is positive semi-definite. Of the greatest importance for QBism is that this statement is an ``if and only if.''

In every exposition of the theorem's importance we know of, it is always presented as an efficient means to check whether a given $\Phi_i$ arises from a valid dynamical account or not. If not, $\Phi_i$ is said to be ``unphysical.''  But since the Choi theorem is an if and only if statement, one could just as well go the other way around:  One could take it as axiomatic that an update map $\Phi_i$ should generate a positive semi-definite Choi matrix, with the Hellwig-Kraus ``origin story'' now reduced to a particular representation of the map and of no greater consequence than that.

In fact, from the QBist point of view, this is the most pleasing resolution.  Within QBism, it was already an axiom that quantum states belong to the positive semi-definite operators.  Similarly it was an axiom that POVM elements be positive semi-definite operators.  In the case of update maps it is a touch more subtle, but one can make in essence the same axiom there too.  In this case the $\Phi_i$ would not themselves be positive semi-definite operators, but rather drawn from a convex set that is isometrically isomorphic~\cite{Barker1984} to the convex set of positive semi-definite matrices.  That is to say, the geometry of the states, the POVM elements, and all update maps would be identical, aside from the appropriate dimensionalities for each.  One would then see that (once renormalized), all three objects are quantum states or ``quantum states in disguise.''\footnote{It may be worth mentioning that one of us introduced the slogan ``a quantum operation is nothing but a quantum state in disguise'' already 21 years ago~\cite[p.\ 603]{Samizdat2}. The idea as expressed at the time was, ``I want to develop the slogan, `A quantum operation is nothing but a quantum state in disguise.' Lately I am taken with the idea that the  
[Choi] representation theorem is one of the deepest statements in all of physics \ldots\ second only to Einstein’s
principle of equivalence!'' However, somehow the author didn't take it as seriously as his initial enthusiasm suggests.  He only genuinely took it seriously with the present work.}
This, we QBists currently take as an axiom of the theory. Most importantly for the present paper, it gives us a means to give a dynamics-less account of the Kraus rules.  A Kraus update scheme is a judgment no more and no less than a quantum state itself is.  Particularly, the state change rule does not logically descend from a physical dynamics.

\section{Reflection}
\label{ReflectionSection}

The quantum formalism provides a QBist agent with criteria on how to update her beliefs following a measurement action. It is important to note that
formulas such as Eq.\ (\ref{eq:kraus}) must not be viewed as a description of how the agent's beliefs necessarily change. Rather, Eq.\ (\ref{eq:kraus}) provides criteria for how the agent {\it should\/} update her quantum state. The Born rule~(\ref{eq:born}) and the state update rule~(\ref{eq:kraus}) are {\it normative\/}, not descriptive. That is, they do not determine how our agent does in fact act but provide her with guidance on how she {\it should\/} act.

The QBist theory of quantum state updating is closely connected to the theory of how a user of personalist Bayesian probability theory should update her degrees of belief. The most well-known update rule is, of course, Bayes's rule, but there exist several important generalizations~\cite{Jeffrey1965,Skyrms1987}. The theory of personalist belief update was unified and given its definitive form in 1984 when van Fraassen derived his {\it reflection principle\/} from a Dutch book argument~\cite{vanFraassen1984,Shafer1983,Goldstein1983}. See Ref.\ \cite{Fuchs2012a} for a thorough discussion of the principle.  In fact, in that paper we applied the reflection principle to the problem of understanding what the notion of a {\it nonselective\/} measurement could possibly mean from a QBist standpoint. It turns out that van Fraassen's reflection principle is also key to a QBist
understanding of dynamics.

Let us first review the (classical) reflection principle and its origins before coming back to the quantum case.

\subsection{Classical Reflection}

The starting point for the reflection principle is the general conception of personalist Bayesian probability as being a normative calculus.  A simple way to give this an operational meaning is through an agent's betting preferences.
When an agent assigns probability $p$ to an event $E$, she regards $\$p$ to be the fair
price of a standard lottery ticket that pays $\$1$ if $E$ is true. In other
words, an agent who assigns probability $p$ to the event $E$ regards both
buying and selling a standard lottery ticket for $\$p$ as fair transactions;
for her, the ticket is worth $\$p$.

A set of probability assignments is called {\it incoherent\/} if it can lead to a sure loss in the following sense: There exists a combination of transactions consisting of buying and/or selling a
finite number of lottery tickets which (i) leads to a sure loss---i.e., a loss no matter what events occur---and (ii) the agent regards as fair according to these probability assignments. A set of probability assignments is {\it coherent\/} if it is not incoherent. A key example of what we have been meaning by a ``normative principle'' is that an agent should aim to avoid incoherent probability assignments.

The famous Dutch-book argument \cite{Ramsey26,DeFinetti90,Jeffrey04}
establishes that an agent's probability assignments $P_0$ at a given time $t=0$ are
coherent if and only if they obey the usual probability rules: For instance, that $0\le
P_0(E)\le1$ for any event $E$, that $P_0(E\vee D)=P_0(E)+P_0(D)$ for mutually exclusive events $E$ and $D$, and so forth.

In this approach, conditional probability is introduced as the fair price
of a lottery ticket that is refunded if the condition turns out to be false.
That is, let $D$ and $E$ be events, and let $\$q$ be the price of a
lottery ticket that pays $\$1$ if both $D$ and $E$ are true, and $\$q$ (thus
refunding the original price) if $D$ is false. For the
agent to make the conditional probability assignment $P_0(E|D)=q$ means that she
regards $\$q$ to be the fair price of this ticket.

It is then a consequence of Dutch-book coherence that the product rule
$P_0(E,D)=P_0(E|D)P_0(D)$ must hold \cite{Jeffrey04}. In other words, conditional probability
assignments violating this rule are incoherent. If $P_0(D)\ne0$, we obtain
 Bayes's rule,
\begin{equation} \label{eq:bayesRule}
P_0(E|D)=\frac{P_0(E,D)}{P_0(D)} \;.
\end{equation}
This shows that a coherent agent must use Bayes's rule to set the
conditional probability $P_0(E|D)$. The value of $P_0(E|D)$ expresses what
ticket prices the agent regards as fair at time $t=0$, i.e., before she finds
out the truth value of either $D$ or $E$. It says nothing about what ticket
prices she will regard as fair at some later time $t>0$.

However now suppose that at a later time $t=\tau$, the agent learns that $D$ is true and updates her probability for $E$ to some $P_\tau(E)$. Usual Bayesian conditioning consists of setting
\begin{equation}
P_\tau(E)=P_0(E|D)\;.
\label{NormaBayes}
\end{equation}
But, need it be so?

It was first pointed out by Hacking \cite{Hacking1967} that there is no
coherence argument that compels the agent to take into account the earlier
probabilities $P_0(E|D)$ when setting the later probabilities $P_\tau(E)$.  That is to say, {\it without further assumptions}, coherence does not
compel the agent, at $t=\tau$, to use the Bayesian conditioning rule (\ref{NormaBayes}). The way coherence arguments connect probability assignments at different times is more subtle. It is expressed elegantly through van Fraassen's reflection principle \cite{vanFraassen1984}, which itself entails the related constraints of Shafer \cite{Shafer1983} and Goldstein \cite{Goldstein1983} which we will use heavily in this paper. The key idea behind the reflection principle is to consider the agent's beliefs about her own future probabilities, i.e., to consider expressions such as
$P_0\big(P_\tau(E)=q\big)$.

To take the simplest possible scenario, suppose that at $t=0$ an agent is certain that, at $t=\tau$, her probability of $E$ will be $q$, with $q\ne P_0(E)$. In probabilistic terms,
\begin{equation}
P_0\Big(P_\tau(E)=q\Big) = 1 \;.
\label{LittleBigHorn}
\end{equation}
In the case $q<P_0(E)$, this means that, at $t=0$, she is willing to buy a ticket for $\$P_0(E)$ although she already knows that later she will be willing to sell it for the lower price $\$q$. In the case $q>P_0(E)$, it means that, at $t=0$, she is willing to sell a ticket for $\$P_0(E)$ although she already knows that later she will be willing to buy the same ticket for the higher price $\$q$. In both cases, already at $t=0$ the agent is certain of a sure loss.  Therefore, Eq.\ (\ref{LittleBigHorn}) requires the agent adjusts her mesh of probabilities so that
\be
P_0(E) = P_\tau(E)\;,
\label{WolfmanJack}
\ee
else she exposes herself to a Dutch book.

It turns out that this simple scenario contains the main idea of van Fraassen's ``diachronic'' Dutch book
argument. Similar to the coherence argument for a single time, we
say that an agent's probability assignments are incoherent if there exists a
combination of transactions consisting of buying and/or selling a finite
number of lottery tickets at two different times such that (i) {\em already at the
earlier time}, the agent is sure of a net loss; and (ii) each transaction is
regarded as fair by the agent according to her probability assignments {\em at the
time the transaction takes place}. As before, an agent should aim to avoid incoherent probability assignments.

To turn the simple scenario into the full-fledged diachronic Dutch-book
argument, one only needs to relax the assumptions that $q \ne P_0(E)$ and that at $t=0$ the agent is
{\it certain\/} that $P_\tau(E)=q$.
Then it can be shown that an agent's probability assignments are incoherent unless
\begin{equation}  \label{eq:reflection}
 P_0\Big(E\,\Big|\,P_\tau(E)=q\Big) = q \;,
\end{equation}
i.e., unless at $t=0$ the agent's conditional probability of $E$, given that
$P_\tau(E)=q$, equals $q$. This is the {\it reflection principle}.

To get this in a form that we will need for much of the remainder of the paper (i.e., in the form of Shafer \cite{Shafer1983} and Goldstein \cite{Goldstein1983}), we just use a simple combination of single-time coherence along with the reflection principle.  Suppose the agent, instead of contemplating a single proposition $Q = [P_\tau(E)=q] $ for what she will believe of $E$ at $t=\tau$, contemplates a range of distinct propositions $\{Q\}$ to which she assigns probabilities $P_0(Q)$.  Then, single-time coherence requires
\begin{equation}
P_0(E)=\sum_Q P_0(Q) P_0(E|Q)\;,
\end{equation}
for which reflection in turn implies
\begin{equation}
P_0(E) = \sum_q P_0\Big(P_\tau(E)=q\Big)\, q\;.
\label{Hermeneutic}
\end{equation}

With this in hand, we are ready to state the quantum mechanical version of these considerations. However let us first reemphasize one thing.  Eq.\ (\ref{eq:reflection}) refers to a ticket price at $t=0$ conditioned on an anticipated ticket price at $t=\tau$.  What if by the time $t=\tau$ rolls around something has so changed the agent's world that none of the values $q$ in the sum~(\ref{Hermeneutic}) match the {\it actual\/} fair price $P_\tau(E)$ she deems by then? Well it just does not matter: The agent has done all that she can do to protect her finances from her standpoint at $t=0$.  Reflection is really all about $t=0$.

\subsection{Quantum Reflection}
\label{AnimalPlanet}

Let us return to the scenario of two consecutive quantum measurements. For concreteness, let us specify three discrete moments in time, $t_0$,
$t_1$, and $t_2$---the two consecutive measurements will take place at $t_1$ and
$t_2$, but all the agent's reasoning will happen at $t_0$. Particularly, we assume the consistency criteria provided by the quantum
formalism to the mesh of beliefs the agent holds at $t_0$. These are: the
POVM $\{E_i\}$ the agent has assigned to the first measurement at $t_1$; the system
state $\rho$ that the agent is committed to for her gambles on the outcomes of
the $\{E_i\}$ measurement; her outcome probabilities $p_i={\rm tr}(\rho E_i)$;
and finally the updated states $\rho_i$ given by Eq.\ (\ref{eq:kraus}) that she
expects to use at $t_2$ if she gets outcome $i$ in the first measurement.

Now assume the agent wants to make a decision at $t_0$ about what measurement
to take at $t_2$. How should she take this into account?  Well, since she has not yet done the $t_1$ measurement, she cannot use
one of the conditional states $\rho_i$\@. She only has a probability $p_i$ that $\rho_i$ will be her
updated state at $t_2$. Of course, a quantum version of the reflection principle comes immediately to mind.

However, one might have a lingering worry that all the elements in classical reflection are probability distributions, whereas quantum states are ostensibly something else, namely positive semi-definite operators on a complex Hilbert space.  In QBism though, that is not a real difference; rather a quantum state should be thought of as {\it exactly\/} a probability distribution: A probability distribution for the outcomes of a chosen informationally complete  ``reference measurement''~\cite{Fuchs13a}. Thus there can be nothing particularly intricate or new in applying the reflection principle to quantum states. Reflection requires that at $t_0$ the agent uses the ``reflected state''
\begin{equation}
  \tilde{\rho}= \sum_i p_i \rho_i = \sum_{i} A_{i}\rho A_{i}^\dagger
  \label{eq:cptp}
\end{equation}
for her decisions about what measurement actions to take at $t_2$, where hereafter we will use a tilde to denote a state obtained via reflection on a planned measurement. For instance, if the agent contemplates
performing the POVM $\{F_j\}$ to the second measurement, then at time $t_0$ her
probability for outcome $j$ should be given by ${\rm tr}(\tilde{\rho} F_j)$.

The state change from $\rho$ to $\tilde{\rho}$ arises because the agent is
intending to perform a measurement action at $t_1$. The agent uses the state $\tilde{\rho}$
if she needs to make a decision at $t_0$ about her actions at $t_2$.  The
reflected state $\tilde{\rho}$ is not, however, the state the agent anticipates to
hold at time $t_2$, following her measurement at $t_1$. Indeed, at $t_2$ she
expects to have obtained one of the outcomes $i$ and to use the conditional
$\rho_i$ for her decisions. Moreover, just as with the classical case, who knows, perhaps something so earthshaking will happen for the agent between $t_0$ and $t_2$ that she will throw away all her previous beliefs (including the $\rho_i$) and start completely afresh. What is important is that agent has done all she can to protect her finances given what she expects at $t_0$.

The map from $\rho$ to $\tilde{\rho}$ defined by Eq.\ (\ref{eq:cptp}) is completely
positive and trace preserving (CPTP). It is a well-known fact that every CPTP
map can be realized in this way for some measurement at $t_1$. It should be
clear, however, that in this section we have not derived that
Eq.\ (\ref{eq:cptp}) applies to quantum dynamics, i.e., state change in the
absence of measurement action. As we stressed in the previous paragraph, our
agent will use $\tilde{\rho}$ only for decisions she makes at $t_0$ and only if she
intends to act on the system at $t_1$. Furthermore, she does not normally
anticipate to adopt $\tilde{\rho}$ at $t_2$, but rather one of the $\rho_i$ instead.

\section{Dynamics}
\label{DynamicsSection}

Let us use the notation $\rho_{\beta|\alpha}$ for a quantum-state assignment an agent makes at time $t_\alpha$ for analyzing any real or hypothetical measurement she could make at a time $t_\beta$. Probabilities $p_{\beta|\alpha}$ for the measurement outcomes are understood similarly. Recall now the setting of Section~\ref{AnimalPlanet} with a $t_0$, $t_1$, and $t_2$\@.  It follows that if our agent contemplates at $t_0$ a measurement action at $t_2$ she actually intends to do, then she will assign a state $\rho_{2|0}$ by which to calculate the measurement's outcomes.  However, if at $t_0$ she contemplates a measurement at $t_1$ which she has no intention of doing, she can nonetheless write down a state $\rho_{1|0}$ to guide any hypothetical she might ask.  Similarly so for $\rho_{0|0}$.  In the development ahead we will effectively equate $\rho_{1|0}$ and $\rho_{0|0}$, working under the assumption that our agent contemplates no state update between $t_0$ and $t_1$.

Now suppose that the agent \emph{does} intend to measure the system as specified by a set of Kraus operators $\{A_j\}$ at the intermediate time $t_1$. The agent expects her post-measurement states to be
\begin{equation}
    \rho^j_{2|1}=\frac{A_j\rho_{1|0}A_j^\dag}{p^j_{1|0}}\quad\text{with probabilities}\quad p^j_{1|0}=\tr\rho_{1|0}E_j\;.
\end{equation}
The reflection principle then requires 
\begin{equation}
    \tilde{\rho}_{2|0}=\sum_jp^j_{1|0}\rho^j_{2|1}=\sum_j A_j\rho_{1|0}A_j^\dag\;.
\end{equation}

To address the question of dynamics, the agent's goal is to consult her mesh of beliefs in order to find some constraint on how $\rho_{0|0}$ and $\rho_{2|0}$ should be related. Note that $\rho_{2|0}$ refers to her beliefs about the consequences of any action she could take after a planned period of inaction. How can she attempt to make sense of this gap? Of course, the number of things she could consult is potentially infinite, but, as quantum theory is about action, it is a good guess that this is where her focus should be trained. For instance, she might ask what she would believe {\it were she to instead\/} plan to take a measurement action during this period, perhaps to elicit updated guidance along the way. This hypothetical helps guide us to the key realization. What our agent should be asking is, are there any measurements between $t_0$ and $t_2$ that she {\it could perform\/} at $t_1$ but would nonetheless deem irrelevant for her $t_0$-gambles? That this should be searched for, we take to be one of the key methodological claims of QBism.

Assume the agent has identified a
measurement, characterized by Kraus operators $\{A_{i}\}$, that she regards as
irrelevant for her analysis at $t_0$ about further measurements at $t_2$. Then it follows that she should use the same state at $t_0$ whether or not she plans to make this measurement at $t_1$. In other words, the indifference judgment implies $\rho_{2|0}=\tilde{\rho}_{2|0}$. Her beliefs spanning a period of inaction are normatively anchored to her beliefs concerning one particular measurement. 

With this, we have almost everything we need for a reconstruction of dynamics. Only one more ingredient is required: If, in addition to the agent's judgment that the measurement is irrelevant to her $t_0$ analysis of measurements she could make at $t_2$, she also deems this to be independent of $\rho_{1|0}$, then she should use the CPTP map
\begin{equation}
    \Phi(\rho):=\sum_j A_j\rho A_j^\dag
\end{equation}
to relate $\rho_{0|0}$ and $\rho_{2|0}$, namely $\rho_{2|0}=\Phi(\rho_{0|0})$.

We have thus arrived at a QBist version of quantum dynamics. Based on an analysis of a hypothetical measurement our agent could perform but chooses not to because of indifference to it, she should use a CPTP map to relate the states $\rho_{0|0}$ and $\rho_{2|0}$. If she cannot identify an appropriate hypothetical
measurement, the quantum formalism gives her no guidance on how to connect her states. In that case, there is no notion of dynamics.

Finally, it is worth noting that if the agent further believes with certainty that she will hold the state $\rho_{2|2}$ at the later time $t_2$, then another invocation of the reflection principle implies that $\rho_{2|2}=\rho_{2|0}$. This follows from the same reasoning as in our discussion surrounding Eqs.\ (\ref{LittleBigHorn}) and (\ref{WolfmanJack}).  In other words, the state the agent expects to hold at the later time will be given by the action of a CPTP map: 
\begin{equation}
    \rho_{2|2}=\Phi(\rho_{0|0})\;,
\end{equation}
which indeed looks like the usual construal of complete positivity in objectivist accounts of it.

\subsection{A Concrete Example}

Before a discussion of some conceptual points surrounding our general result---for that, see Section \ref{WithoutApologetics}---let us consider a concrete example: The bit-flip channel that appears in every discussion of quantum error correction.

Nielsen and Chuang start their discussion of quantum error correction by
considering a single qubit and an error process that consists of a ``bit flip'' with
probability $\epsilon$~\cite[p.\ 426]{Nielsen2010}. The so-called bit flip corresponds to applying the Pauli $X$
operator to the computational basis $\{|0\rangle,|1\rangle\}$: 
\be
X|0\rangle=|1\rangle \quad \mbox{and} \quad X|1\rangle=|0\rangle\;.
\ee
The basic error process is thus presented as 
\begin{equation}
\rho \quad\longrightarrow \quad X\rho X \quad\; \mbox{with probability}\;\epsilon\;.
\end{equation}

But one should beware:  Such a description corresponds to hybrid of classical and quantum thinking that is immediately suspicious to a QBist mindset.  Whose probability is $\epsilon$ after all?  And what is the meaning of the dynamical process $\rho \,\rightarrow\, X\rho X$ in the
absence of any measurement action?  A.k.a., the very subject of this paper!

\begin{figure}[t]
    \centering
    \includegraphics[width=0.5\textwidth]{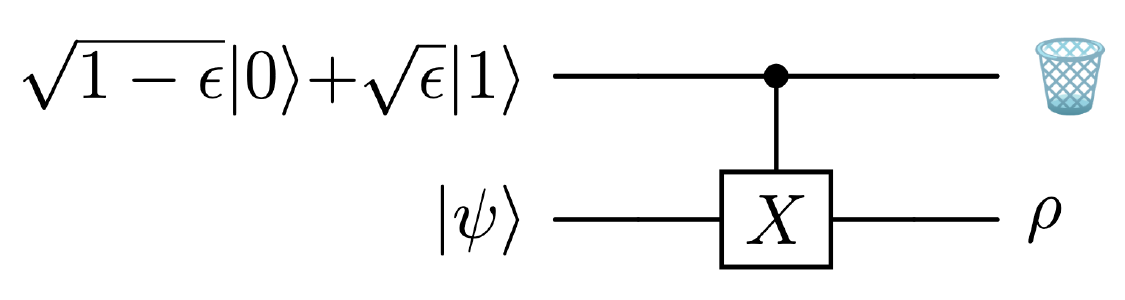}
    \caption{A common way of thinking of a bit-flip channel: a delicately precise initial state assignment for an environment, followed by a precise unitary interaction, and then a discarding of the environment from further consideration.}
    \label{GittyUp409}
\end{figure}

A seemingly more sophisticated account (essentially Everettian) is depicted in Figure~\ref{GittyUp409}, where now the bit-flip process is viewed as a global unitary evolution that entangles the system with an environment.  Of course, the environment must start with a carefully specified initial state, and a QBist asks, ``Whose state assignment is that? Indeed, who has drawn the circuit diagram with its explicit judgment of a unitary interaction?''

The QBist account developed here instead recognizes a decision making agent to begin with. The whole story is about an agent's judgments.  Referring to our previous general discussion, we imagine an agent who considers a potential measurement on the qubit defined by the Kraus operators
\be
A_1=\sqrt{1-\epsilon}\,I \quad\mbox{and}\quad A_2=\sqrt\epsilon\, X\;.
\ee
If for any reason the agent is indifferent as to whether she intends to perform this
measurement at time $t_1$ or not, her state assignments at $t_0$, namely $\rho_{0|0}$ and
$\rho_{2|0}$ should be 
related by the CPTP map
\begin{eqnarray}
\rho_{0|0} \quad \longrightarrow \quad \rho_{2|0} &=& (1-\epsilon)I\rho_{0|0}I + \epsilon X\rho_{0|0}X
\\
&=&
(1-\epsilon)\rho_{0|0} + \epsilon X\rho_{0|0}X\;.
\end{eqnarray}
This is the standard bit-flip channel, but now derived from an assumption of indifference.  

What is distinct technically in our analysis is that it forces the agent to think through which hypothetical measurements at  intermediate time $t_1$ are not worth bothering about with regard to her gambles at $t_0$.  If she can identify one, then that is what she calls the ``error.''  With that understanding, what does error reduction as a strategy amount too?  That the agent should strive to be in a situation where she is less indifferent to any measurement at $t_1$ than the identity measurement $I$ itself.

Other standard qubit channels include the dephasing channel, the depolarizing
channel, and the amplitude-damping channel. All three are very similar to
the bit-flip channel given above and can be treated in exactly the same
way. We will present a continuous-dynamics version of the amplitude damping channel in Section~\ref{ContinuousDynamicsSection} below.

\subsection{Discussion}
\label{WithoutApologetics}

Some remarks about our assumptions are in order. To a first glimpse it might appear strange that an agent would intentionally choose not to perform the very measurement she had contemplated for working out her $t_0$-beliefs about her $t_2$-gambles.  For instance, it is well known that the von Neumann entropy of $\tilde{\rho}$ is generally larger than the entropies for the individual $\rho_i$~\cite{FuchsJacobs}:
\be
S(\tilde{\rho})\ge\sum_i p_i S(\rho_i)\;.
\ee
Thus, with a measurement, the agent would have more predictability to make use of in her real-time gambles at $t_2$. However, one must remember that diachronic coherence is a game of consistency concerning one's beliefs {\it purely\/} at $t_0$.  And one of our agent's beliefs is that she will not perform a measurement at $t_1$---so everything has to fall in place around that.

Perhaps an example drawn from the double-slit experiment will be useful to see our key conceptual points.  Imagine an experiment where {\it no\/} interference pattern is expected to be produced. For instance, perhaps the double-slit experiment is being done with marbles, where no one has ever seen an interference pattern before. In the common decoherence account, one says that the pattern is ``as if'' there were a measurement that determines which slit the particles went through. Basically, the usual intuition is that there is an actual measurement going on (with a measurement record in the environment) which {\it causes\/} the interference pattern to disappear.

QBists replace this causal account by a consistency criterion. Instead of the above ``as if'' description, our criterion is that we expect the same pattern whether or not we make the which-slit measurement. Instead of talking about the measurement causing the disappearance of the pattern, we insist that our three judgments---$\rho$ at $t_0$, odds at $t_0$ for gambles at $t_2$, and measurement at $t_1$ irrelevant for the latter---are all on an equal footing.  That obviates the need to talk about an environment at all.

Furthermore, even if there were such an environment, just exactly what would it be measuring?  From Choi's theorem~\cite[p.\ 372]{Nielsen2010} it is known that if the Kraus operators $A_i$ are good for defining the CPTP map in Eq.\ (\ref{eq:cptp}), then so too will be any Kraus operators of the form
\be
B_j=\sum_i u_{ji} A_i
\ee
where $u_{ji}$ are the matrix elements of any isometry.  So there is no single measurement whose outcomes the environment could record.  A QBist would say, ``Of course, because there is no measurement at $t_1$ anyway. A measurement requires the action of an agent, and there is none to be found in the environment.''  For a QBist contemplating a hypothetical measurement, it is of no consequence that there might exist another measurement leading her to adopt the same CPTP map.

Finally let us discuss a delicate point.  QBism is often judged by philosophers of physics to go to extremes.  For instance, QBism insists that probability-1 (i.e., certainty) be treated as a personalist degree of belief or gambling commitment just like any other probability.  Here we expose a new stance---one that might also be judged extreme---to accompany our previous ones. In Section~\ref{QBismSection}, we stated that all actions an agent can take on a physical system correspond to the POVMs $\{E_i\}$ appropriate to the system's Hilbert space. Well then, the POVM $\{I\}$ with a single element must be an action too.

But what could it mean for the identity operator $I$ to be an action? A hint of an answer can be found in another statement of Section~\ref{QBismSection}: ``When the consequences of the actions matter to the agent in the sense that they are worth her while to gamble upon them, she is well advised to use the quantum formalism to guide her decisions regarding which actions to take.'' To identify a system as one that matters for the agent's gambling is, in effect, to take an action on it. One has to conceptually carve it out of its background and ask how best to gamble on later, further actions upon it.  Among other things, $I$ defines a Hilbert space dimension for the system.  More particularly, the identity operator $I$ corresponds to the action whose consequence is the sense of the object's existence for the agent.  So long as the agent writes down any quantum state $\rho$ at all, she will expect with certainty that her sensed experience will be a ``yes.'' For trivially, ${\rm tr} (\rho I) ={\rm tr}\,\rho=1$.  Still, taking the action $I$ is not the same thing as ``doing nothing.''

Thus for the case of the POVM $\{I\}$, as for all others in Eqs.\ (\ref{ChurchBells})--(\ref{WillBeRinging}), or the less general case Eqs.\ (\ref{BoogalooSituation})--(\ref{eq:kraus}), an agent should use our representation theorem for defining a quantum dynamics when she deems taking the action $I$ or not at $t_1$ irrelevant for her $t_0$ gambles. The agent must then use a Kraus decomposition of $I$ for her reflection.

If that Kraus decomposition only has one element, then $A^\dagger A=I$ and so $A$ must be unitary.  But as before, the unitary $A$ is something that only lives on the agent's spreadsheet at time $t_0$. It is not something that exists out in the world, as it would be for instance in an Everettian interpretation of quantum mechanics.

Overall, we see that our criterion treats both so-called ``closed-system'' and ``open-system'' dynamics on an equal footing.  It is just that what we mean by a ``closed system'' is that the measurement reflected upon at $t_1$ is the singleton set $\{I\}$. As we have seen, this can lead to a unitary quantum dynamics. But there are more possibilities in our meaning of closed system than can be found in the standard view. Particularly per Eq.~(\ref{ChurchBells}), any Kraus decomposition of the identity
\begin{equation}
I = \sum_k A_k^\dagger A_k
\end{equation}
can also be a closed-system dynamics when used in our representation theorem.  This indeed hearkens back to Asher Peres's observation with regard to black-hole physics that there is no reason to believe that all CPTPs must be derivable from an overall unitary dynamics~\cite{Peres2004}.

A key point is that our derivation of the CPTP maps describing quantum dynamics does not invoke an environment.\footnote{This is not to say that there is no environment for the system. It is just that the environment need not be part of the agent's concern.} It is entirely couched in the agent's beliefs about the consequences of measurement actions she might take on the system itself.  Nor does our derivation rely on the notion of a ``nonselective measurement'' that goes back all the way to von Neumann. That is, a measurement that genuinely takes place but without its outcome being learned by an agent. ``Nonselective measurement'' is an oxymoron in QBism. This is because in QBism a measurement is fundamentally an action an agent takes on her external world precisely to elicit a new experience for herself.

\section{Continuous Dynamics}
\label{ContinuousDynamicsSection}

The CPTP map~(\ref{eq:cptp}) describes a discrete quantum channel.  
To arrive at continuous dynamics,
the agent will need to contemplate not just a single hypothetical measurement
at an intermediate time $t_1$, but a continuous measurement in the time
interval $[t_0,t_2]$.

The theory of continuous measurements is a straightforward generalization of
the Kraus-operator formalism of the previous sections. Our exposition follows
closely Ivan Deutsch's lecture notes on quantum trajectories~\cite{Deutsch2023}. Assume an agent
is making a continuous measurement on a quantum system, and that her state at
time $t$ is $\rho(t)$. For simplicity, we only consider measurements
with a binary outcome: At any given time, the agent either does or does not
register a ``click.'' We also assume her Kraus operators do not depend on previous measurement outcomes, which amounts to a Markov assumption. Leaving the current time index implicit, the Kraus operator corresponding to a click
in the interval $[t,t+dt]$ is of the form
\begin{equation}
  M_1(dt)=L\sqrt{dt} \;,
\end{equation}
where the so-called Lindblad operator $L$ is an arbitrary linear map. The
probability of a click in the interval $[t,t+dt]$ is then
\begin{equation}
  dp(t) = {\rm tr}\big(M_1^\dagger(dt) M_1(dt) \rho(t)\big) = {\rm tr}\big(L^\dagger
  L\rho(t)\big)\,dt\;
\end{equation}
and the agent's post-measurement state after registering a click is
\begin{equation}
  \rho(t+dt) = \frac{M_1(dt) \rho(t)M_1^\dagger(dt) }{dp(t)}
  = \frac{L\rho(t)L^\dagger}{{\rm tr}\big(L^\dagger  L\rho(t)\big)}\;.
\end{equation}
The Kraus operator $M_0(dt)$ corresponding to no click in the interval $[t,t+dt]$ satisfies
\begin{equation}
  M_0^\dagger(dt)M_0(dt)=I-M_1^\dagger(dt)M_1(dt)=I-L^\dagger L\, dt\;.
\end{equation}
The general form of $M_0(dt)$ is, in polar decomposition,
\begin{equation}
  M_0(dt)=\sqrt{I-L^\dagger L\, dt}\,\exp\!\left(-\frac{i}{\hbar}H\,dt\right)
\end{equation}
for some hermitian operator $H$ (the Hamiltonian).
Hence, neglecting higher-order terms in $dt$,
\begin{equation}
M_0(dt)=\left(I-\frac12 L^\dagger L\, dt\right)\!\left(I-\frac{i}{\hbar}H\,dt\right)
=I-\frac12 L^\dagger L\, dt-\frac{i}{\hbar}H\,dt\;.
\end{equation}
The agent's probability for not registering a click in the time interval
$[t,t+dt]$ is
\begin{equation}
{\rm tr}\big(M_0(dt)^\dagger M_0(dt) \rho(t)\big) = 1-dp(t)
 \end{equation}
and her post-measurement state in this case is
\begin{equation}
  \rho(t+dt)=\frac{M_0(dt)\rho(t) M^\dagger_0(dt)}{1-dp(t)}
  = \frac{\Big(I-\frac12 L^\dagger L dt-\frac{i}{\hbar}H\,dt\Big)\, \rho(t)\,
  \Big(I-\frac12 L^\dagger L dt+\frac{i}{\hbar}H\,dt\Big)}{1-dp(t)}\;,
\end{equation}
which is equivalent to the differential equation
\begin{equation}
  \frac{d\rho}{dt} = -\frac i\hbar [H,\rho]  -\frac12
  \big(L^\dagger L\rho+ \rho L^\dagger L\big) \;.
\end{equation}

To make the following discussion more concrete, consider a two-level atom driven by a resonant classical laser
field and coupled to the vacuum. This is one of the simplest physical examples
from quantum optics. Again, our discussion follows Deutsch's
notes~\cite{Deutsch2023} closely. An agent who has such a system in front of herself can consider
monitoring it continuously using a photodetector. In other words, she may
consider to perform a continuous quantum measurement where, at any moment, she
gets one of two outcomes: The photodetector either does or does not register a
photon.

The Kraus operators $M_1(dt)$ (a click) and $M_0(dt)$ (no click) quantify the
agent's beliefs about the consequences of her measurement action. These
beliefs will be informed by a large number of factors, including the agent's
education in physics and quantum optics, her experience as an experimenter, the
results of any calibration measurements she did when setting up the experiment,
her judgment to use the rotating wave approximation in this case, etc.

As we have seen, assigning the Kraus operators
$M_1(dt)$ and $M_0(dt)$ is equivalent to choosing a Lindblad operator $L$ and a
Hamiltonian $H$.  We will assume that the above considerations have led the
agent to the somewhat idealized but standard choices
\begin{equation}
L = \sqrt{\Gamma}\, \sigma_- \;\mbox{ and }\;
H=-\frac{\hbar\Omega}{2} \big(\sigma_+ \, +\, \sigma_-\big)\;,
\end{equation}
where $\Gamma$ and $\Omega$ are the decay rate and the field coupling constants, and the lowering
and raising operators $\sigma_-=|g\rangle\langle e|$ and
$\sigma_+=|e\rangle\langle g|$ are defined in terms of the ground and excited states
$|g\rangle$ and $|e\rangle$ of the two-level atom.

At any time, the agent's probability for registering a photon in the time interval
$[t,t+dt]$ is given by
\begin{equation}
dp(t) = {\rm tr}\big(L^\dagger L \rho(t)\big) = \Gamma \,{\rm tr}\big(\sigma_+\sigma_-
  \rho(t)\big)\,dt = \Gamma \langle e|\rho(t)|e \rangle\, dt \;. \label{eq:dp}
 \end{equation}
Upon detecting a photon, her post-measurement state is
\begin{equation}
  \rho(t+dt)=\frac{\Gamma\sigma_-\rho(t)\sigma_+ dt}{dp(t)}
  = \frac{\Gamma dt}{dp(t)}|g\rangle\langle e|\rho(t)|e\rangle\langle g| =
  |g\rangle\langle g|\;,
\end{equation}
and if she doesn't detect a photon, she gets
the differential equation
\begin{equation}
  \frac{d\rho}{dt} = -\frac i\hbar [H,\rho]  -\frac\Gamma 2
  \big(\sigma_+\sigma_-\rho+ \rho\sigma_+\sigma_-\big) \;.  \label{eq:de}
\end{equation}
The solution $\rho(t)$ of this differential equation allows her to evaluate her
probability for not detecting a photon in a finite time interval $[t,t+\tau]$,
\begin{equation}
p\Big(\text{no click} \in [t,t+\tau]\Big) = 1-\int_t^{t+\tau} dp(t) = 1-\int_t^{t+\tau} \Gamma
\langle e|\rho(t)|e \rangle\, dt \;. \label{eq:p}
\end{equation}

Let us take stock---note particularly how we have presented all this. At time $t=t_0$, the agent's quantum state for the two-level
atom is $\rho(t_0)$. She uses this state to guide her decisions about gambles at
$t=t_0$. She now contemplates a continuous measurement lasting from $t_0$ to
$t_2$, during which she expects to detect a number $n\ge0$ of photons. Starting
from her initial state $\rho(t_0)$, she can use equations~(\ref{eq:dp}),
(\ref{eq:de}) and (\ref{eq:p}) to compute her probability density
$p(n;\tau_1,\ldots,\tau_n)$ of detecting exactly $n$ photons at the times
$\tau_1,\ldots,\tau_n\in[t_0,t_2]$. Furthermore, she can write down the
conditional state $\rho(n;\tau_1,\ldots,\tau_n)$, which informs
her decisions at $t=t_0$ about conditional gambles at $t=t_2$, i.e., gambles
conditional on her detecting exactly $n$ photons at the times
$\tau_1,\ldots,\tau_n\in[t_0,t_2]$.

If the agent now averages over all possible measurement outcomes, she obtains
the reflected state
\begin{equation}
  \tilde{\rho}(t_2) = \langle \rho(n;\tau_1,\ldots,\tau_n)\rangle_{p(n;\tau_1,\ldots,\tau_n)} \;,
  \end{equation}
  where the notation $\langle \ldots\rangle_{p}$ denotes taking the expectation value with respect to the probability density $p$.
According to the reflection principle, $\tilde{\rho}(t_2)$ informs her
decisions at $t=t_0$ about gambles at $t=t_2$, under the assumption that she
will perform the continuous measurement between $t=t_0$ and $t=t_2$. The
reflected state $\tilde{\rho}(t_2)$ can also be obtained by solving the
Lindblad master equation
\begin{equation}
  \frac{d\rho}{dt} = -\frac i\hbar [H,\rho]  -\frac\Gamma 2
  (\sigma_+\sigma_-\rho+ \rho\sigma_+\sigma_-)
  +\Gamma \sigma_-\rho\sigma_+   \;.  \label{eq:master}
\end{equation}
So far we have not talked about dynamics. All state
changes are the consequence of the agent's planned continuous measurement
actions. To arrive at dynamics requires an additional judgment on the agent's
part: the judgment that, as far as her expectations at $t_0$ about measurements at
$t_2$ are concerned, it does not matter whether she intends to carry out the
continuous measurement or not.

In analogy to her choice of states, Hamiltonians, and Lindblad operators, this
judgment is informed by her striving for consistency, taking into account her
physics education, her experience, other judgments, etc. The above judgment---that whether or not she intends to carry out the continuous measurement is
irrelevant for her expectations at $t_0$ about measurements at $t_2$---is equivalent
to the judgment of continuous dynamics, i.e., the judgment that even in the
absence of any measurement action, the agent should use the solution of the
Lindblad master equation~(\ref{eq:master}) to inform her gambles at $t_2$.

One might ask, why would an agent go to the trouble of doing an intricate calculation such as the one above with its supposition of a continuous measurement between $t_0$ and $t_2$, when in fact she has no intention of doing the measurement? For one thing, good continuous photodetectors are expensive and might require more ancillary equipment, computing power, etc., than she can afford.  The point is, it's in her best interest to seek any guidance she can for how to gamble on measurements at $t_2$, and this is one way to get some. If the agent decides it is a good strategy for her personally, then our method ensures it cannot be broken by a Dutch book at $t_0$.  But it also gives a point of focus for the agent who desires to reduce the decoherence she finds in the situation:  It is the parameter $\Gamma$, the decay rate the agent feels compelled to ascribe to the setup.  If she wants $\Gamma$ reduced, she should search her soul for what might make that happen.  Perhaps it is something so simple as installing more reflective mirrors appropriately.  In any case, even in this highly idealized situation, our approach has the potential to assist an agent in how to design a better experiment.

\section{Decoherence}
\label{DecoherenceSection}

Decoherence occurs when a system's dynamics maps pure quantum states to mixed states, i.e., it causes states to lose their purity. This is often an undesirable property. On the other hand, since decoherence is a function of the dynamics, everything we have said about dynamics will apply to it as well: The QBist conception of decoherence is thus sure to be radically different from the mainstream approach. Particularly, in QBism decoherence is not something that \emph{happens} to a system. It is an aspect of the way an agent connects her beliefs across time. Just like quantum dynamics, decoherence happens only on paper.

It follows that decoherence plays no fundamental role in a QBist understanding of quantum mechanics.  This contrasts with the usual Everettian story~\cite{Wallace2012}, Zurek's einselection program~\cite{Zurek2022}, Healey's quantum pragmatism~\cite{Healey2017}, Zwirn's convivial solipsism~\cite{Zwirn2016}, and Rovelli's relational quantum mechanics~\cite{Adlam2022}.\footnote{Decoherence is sometimes invoked to explain the emergence of the classical world.  Two of us have addressed this idea in a previous paper~\cite{Fuchs2012a}. Key to it is an objective, agent-independent unitary evolution which is explicitly denied in QBism. For a QBist, decoherence cannot be thought of as a \emph{mechanism} that agent-independently ensures a classical world. Nevertheless, it is still a part of the formalism which may connect to an agent's belief that they will encounter ``classicality'' in some subset of their beliefs. Indeed, we expect some elements of the quantum-to-classical program, once properly translated, to provide valuable insights toward just such a goal. However, that is a subject for future work.} Nonetheless, the study of decoherence is still very important, as it is certainly a hurdle to overcome in practical applications of quantum mechanics. For instance, in a quantum computer, decoherence leads to errors which require the costly implementation of quantum error correction. So, it behooves us to compare the conventional interpretation of decoherence to QBism's own.

In the conventional view, decoherence is considered to be caused by the interaction of a system with its environment~\cite{Schlosshauer2007}. As a starter, a {\it mathematically\/} equivalent, though conceptually distinct, version of this can be understood from the QBist perspective as well. Assume an agent has initially assigned a joint quantum state
to a physical system and its environment of the form
\begin{equation}
  \rho_{\rm joint} = \rho \otimes \rho_{\rm env}
\end{equation}
and judges that the joint system undergoes unitary dynamics---we have
seen above how a QBist agent might arrive at this judgment---leading to the later state
\begin{equation}
  \rho'_{\rm joint} = U\big(\rho \otimes \rho_{\rm env}\big)U^\dagger\;,
\end{equation}
where $U$ is unitary. The agent's later state $\rho'$ for the system alone is given by
tracing out the environment,
\begin{equation}
  \rho' = {\rm tr_{\rm env}}\big[ U\big(\rho \otimes \rho_{\rm env}\big)U^\dagger\big].
\end{equation}
From this, it follows that $\rho'$ can be represented as arising from a CPTP map applied to
$\rho$, 
\begin{equation}
  \rho'= \sum_{i} A_{i}\rho A_{i}^\dagger
  \label{eq:cptp2}
\end{equation}
for some set of Kraus operators $\{A_{i}\}$ satisfying $\sum_{i} A_{i}^\dagger
A_{i}=I$. In general, Eq.~(\ref{eq:cptp2}) does not preserve pure states and
hence leads to decoherence.

Note that in principle there is nothing wrong with this derivation from a QBist perspective. Every element of the story, as it was related above, consisted in a personal judgment; no symbol was interpreted ontically.  However, it required the agent to assign a unitary $U$ to the joint system-environment dynamics. In practice an agent rarely if ever possesses such extraordinarily precise beliefs about an environment as to warrant the assignment of a unitary---it is hard enough to muster numerical beliefs about the system alone. Worse still, the agent may not even know what to count as a relevant environment upon which to write a state $\rho_{\rm env}$, much less an interaction $U$.\footnote{A really exquisite example of how difficult it can be to assign a relevant environment for biomolecule interference experiments can be found in Ref.\ \cite{Arndt2017}, where the immensity of the effort is quite manifest. In the end, the authors are left with the statement that the source of their decoherence is ``likely'' residual charges from the focused ion beam milling process that remain in the gratings.}  Thus, the standard derivation of decoherence as sketched above is essentially never available to a QBist in a real-world situation.

As such, it is better to ground the QBist understanding of decoherence on the agent's beliefs about the system itself and her judgments concerning the measurements she might perform on it. This follows from our derivation of quantum dynamics and the fact that decoherence is a property of a quantum dynamical map. Decoherence in a time interval $[t_0,t_2]$ thus normally occurs for an agent if there is a nontrivial intermediate measurement upon the system (not its environment) that the agent regards as irrelevant for her decisions at $t_0$ about her gambles at $t_2$.

Yet, is this way of telling the story of decoherence of any practical benefit? Is it not \emph{useful} to think of decoherence in terms of the environment's access to a system? Indeed, this is how the mainstream story goes: In order to reduce the effects of decoherence, an experimenter should try to  carefully rig a situation where the system is as isolated as she can get it from the environment. How can a QBist limit decoherence without defaulting to the conventional language and metaphors?

The key, as it has been throughout our development, is that everything begins with the agent's beliefs and the demand that they be Dutch-book coherent with each other and with the Born rule~\cite{DeBrota2021}. Recall the marble example we gave in Section \ref{WithoutApologetics}. The starting point is the agent's expectations for marbles and the kinds of things she thinks she can do with them. Present technological capabilities render it unimaginable that she should \emph{ever} care if she planned to ``watch'' a marble in motion as it approaches the slits. This indifference leads her to assign a purely classical (decoherent) dynamics per our general argument. In order to gamble otherwise---that is, to believe that the decoherence has been lessened---she would need to believe that a measurement \emph{actually would} make a difference for her gambles.  

But, how can she get to that belief?  An agent cannot just choose what she believes.  Marcus Appleby once put it forcefully, ``It's really difficult to believe something you don't actually believe''~  At best the agent can take actions she hopes will result in directing her future beliefs toward a desired target---for instance, toward a purity-preserving dynamics.  So, it is about the agent's actions in the laboratory.

Predicting meaningfully coherent dynamics for a marble is certainly far-fetched for now, but three decades ago most experimenters would have said the same thing even about a shockingly smaller physical system, the fullerene, with 22 orders of magnitude less atoms! In 1999 however, Arndt et al.\ \cite{Arndt1999} managed to observe an interference pattern in a double-slit experiment performed with buckyballs, soccer-ball shaped molecules composed of 60 carbon atoms. Rather than explain what allowed for this monumental achievement as the process of shielding the systems from the prying eyes of the environment, we can simply recognize that the experimenters managed to produce a situation in which they {\it could not help but feel\/} essentially {\it different\/} about an alternative one in which they could ``monitor the flights'' of the buckyballs. 

Still, who could blame the meticulous scientist working on a sensitive experiment in their lab for believing that this requires elaborate shielding {\it from something}? There is truth to this, but the language often used for the intuition is backward from the perspective of QBism.  Take this passage from the Arndt et al.\ paper~\cite{Arndt1999}:
\bq
In quantum interference experiments, coherent superposition
only arises if no information whatsoever can be obtained, even in
principle, about which path the interfering particle took. Interaction with the environment could therefore lead to decoherence. \ldots

In an experiment of the kind reported here, `which-path' information could be given by the molecules in scattering or emission processes, resulting in entanglement with the environment and a loss of interference. Among all possible processes, the following are the most relevant: decay of vibrational excitations via emission of infrared radiation, emission or absorption of thermal blackbody radiation over a continuous spectrum, Rayleigh scattering, and collisions.
\eq
Thus one might think all these environments are crucial to the analysis.  However, to say things in this way loads the dice for one to think there is a path genuinely taken by the buckyball, with or without a measurement on the environment, which one might gain information about through the environment's record.  To the extent that such a description is conceptually incoherent---as QBism believes it is and as was argued in detail in Sections~\ref{OhWolfgangWolfgang} and \ref{WithoutApologetics}---{\it it cannot help the understanding of the experiment or even its design\/} to retain this kind of terminology.  This is just simple logic.  We thus are led to answer the question of the practical benefit of the QBist conception of decoherence with another question:  How much further might experimentalists of the future go if they have learned to think in the QBist way?

We wish we had the prescience to be able to answer this, but the method forward is clear: Adopt the conception and see where it goes. The underlying story must ultimately be one of experimental practice, probabilistic data analysis, and lots and lots of tweaking on the lab bench. We have learned enough from insisting on a Bayesian understanding of quantum probabilities to know that sometimes the technical fruits of the conception can be profound. For instance, even ascertaining the expectation that different trials of the ``same'' experiment are sufficiently the same is not at all a trivial\footnote{In the terminology of Bayesianism, the experimenter must secure an ``exchangeable prior'' for the experimental runs, meaning that they have come to believe that the order of their data is irrelevant, and that this would be true regardless of how much data they plan to collect.} and the solution to the problem leads to both the classical and quantum de Finetti representation theorems~\cite{Bernardo1994,Caves2002b}. One might have initially thought them to be inconsequential philosophical statements, but instead they have been deployed to analyze a rather large number of problems in quantum information science. See \cite{Renner2008,Doherty2004,Renner2021,Enk2002} for a small but relevant sampling. We expect the same here after the passage of time.

Most fundamentally, what we have identified is that the conventional account of decoherence tries to give a God's eye view of what is happening in the laboratory, while no living experimentalist can have as much. QBism is thus tailor-made for the real-world situation.  Beyond this, QBism gives the recognition that ``decoherence'' has no magic ingredient that makes it more fundamental to the story of quantum mechanics than the simple issue of why there is such a thing as quantum dynamics at all.  As previously emphasized, every problem we highlighted as plaguing the conventional accounts of quantum dynamics plagues decoherence as well. Thus it is hard to see how QBism's conceptually consistent story could {\it not\/} be a rightful guide to experimental practice.

\section{Conclusion}
\label{ConclusionsSection}

QBism dissolves the measurement problem by realizing that quantum theory should be interpreted as a decision theory concerning the actions and consequences of any agent wishing to better navigate her world. There is no measurement problem because quantum theory is about action. But in doing so, QBism creates the dynamics problem: How should an agent update her quantum state in times of \emph{inaction}, when nothing is happening to her? Why would she believe her state at a future time should be evolved by a unitary operator or subject to some particular completely positive map? In this paper we have answered this question. True to QBist principles, we find that constraints on inaction ultimately derive from constraints on action. In other words, an agent's beliefs about dynamics derive from her beliefs about measurements.

Specifically, we have shown that the traditional form of quantum dynamics follows normatively from the judgment that there is a measurement an agent could perform some time in the future which she regards as irrelevant for her current gambles about a farther future. This representation theorem resolves the paradox of quantum dynamics similarly to how the quantum de Finetti theorem explains what an ``unknown quantum state'' is from a QBist view. It means that all dynamics, from a spin precessing in a magnetic field to the channel for a circuit implementing Shor's algorithm, are descriptions of belief update fundamentally grounded in beliefs about measurement actions. In direct contrast to approaches which have tried to view observers and observations as derivative of the fundamental dynamics of the universe, QBism tells the reverse story: Measurement is the fundamental thing---all the rest follows from consistency.

Our treatment is notable for its conceptual parsimony. No environment or larger Hilbert space is ever explicitly required; instead, true to its decision theoretic character, only the agent-system binary is needed. Similarly, unitary and non-unitary dynamics are found to be on an equal footing---both arise from the same kind of judgment, only differing in the identity of the measurement judged to be irrelevant. Finally, the concept of measurement itself is never strained to include nonselective ones. Dynamics follows without ever weakening the clean formulation of measurement as an action-consequence pair for an eliciting agent.

As it is a property of a dynamical map, the phenomenon of decoherence obtains a straightforward yet quite unorthodox retelling in our story. The environment and entanglement, the usual suspects, are now supporting actors rather than the responsible parties: A QBist never needs to invoke a second system or an entangling unitary. QBism does not view decoherence as the cause of non-unitary dynamics, but as a simple consequence of it.

Finally, to come back to our story of John Wheeler’s influence, QBism is a step closer to finding its  ontological statement.

\section*{Acknowledgments}

We thank Howard Barnum, Hans von Baeyer, Philipp Berghofer, Michel Bitbol, Ivan Deutsch, Paolo Perinotti, Jacques Pienaar, Tzula Propp, Blake Stacey, and Bill Wootters for discussions and an anonymous referee for pushing us to improve the paper. CAF particularly thanks Harvey Brown for the motivating question of this essay and Matthew Weiss for insights on completely positive maps. JBD acknowledges support from the National Science Foundation Grant PHY-2116246. CAF and RS acknowledge support from the John Templeton Foundation through Grant 62424. The opinions expressed in this publication are those of the authors and do not necessarily reflect the views of the John Templeton Foundation. CAF further acknowledges partial support for this work under National Science Foundation Grants 2210495 and 2328774.

\end{document}